\documentclass[]{emulateapj}
\usepackage{psfig}
\usepackage{amsmath}

\newcommand{\maxi}{MAXI J0158-744}
\newcommand{\oviii}{O \textsc{viii}}
\newcommand{\hei}{He \textsc{i}}
\newcommand{\heii}{He \textsc{ii}}

\newcommand{\lum}{\,erg\,s$^{-1}$}
\newcommand{\kms}{\,km\,s$^{-1}$}
\newcommand{\cm}{\,cm$^{-2}$}
\newcommand{\nh}{$N_\mathrm{H}$}
\newcommand\T{\rule{0pt}{2.6ex}}
\newcommand\B{\rule[-1.2ex]{0pt}{0pt}}
\shorttitle{A Luminous Be+White Dwarf Supersoft Source in the Wing of the SMC: \maxi}
\shortauthors{Li et al.}

\begin{document}
\title{A Luminous Be+White Dwarf Supersoft Source in the Wing of the SMC: \maxi}
\author{
K. L. Li\altaffilmark{1}, 
Albert K. H. Kong\altaffilmark{1, 8}, 
P.A. Charles\altaffilmark{2}, 
Ting-Ni Lu\altaffilmark{1}, 
E.S. Bartlett\altaffilmark{2}, 
M.J. Coe\altaffilmark{2}, 
V. McBride\altaffilmark{3, 4}, 
A. Rajoelimanana\altaffilmark{3, 4}, 
A. Udalski\altaffilmark{5}, 
N. Masetti\altaffilmark{6}, 
Thomas Franzen\altaffilmark{7}
}

\altaffiltext{1}{Institute of Astronomy and Department of Physics, National Tsing Hua University, Taiwan}
\altaffiltext{2}{School of Physics and Astronomy, University of Southampton, Highfield, Southampton SO17 1BJ, UK}
\altaffiltext{3}{South African Astronomical Observatory, P.O. Box 9, Observatory 7935, South Africa}
\altaffiltext{4}{Department of Astronomy, University of Cape Town, Private Bag X3, Rondebosch 7701, South Africa}
\altaffiltext{5}{Warsaw University Observatory, Aleje Ujazdowskie 4, 00-478 Warsaw, Poland}
\altaffiltext{6}{INAF - Istituto di Astrofisica Spaziale e Fisica Cosmica di Bologna, via Gobetti 101, I-40129 Bologna, Italy}
\altaffiltext{7}{CSIRO Astronomy and Space Science, PO Box 76, Epping NSW 1710, Australia}
\altaffiltext{8}{Kenda Foundation Golden Jade Fellow. }

\begin{abstract}
We present a multi-wavelength analysis of the very fast X-ray transient \maxi, which was detected by MAXI/GSC on 2011 November 11. The subsequent exponential decline of the X-ray flux was followed with \textit{Swift} observations, all of which revealed spectra with low temperatures ($\sim$100eV) indicating that \maxi\ is a new Supersoft Source (SSS). The \textit{Swift} X-ray spectra near maximum show features around 0.8 keV that we interpret as possible absorption from \oviii, and emission from O, Fe, and Ne lines. We obtained SAAO and ESO optical spectra of the counterpart early in the outburst and several weeks later. The early spectrum is dominated by strong Balmer and \hei\ emission, together with weaker \heii\ emission. The later spectrum reveals absorption features that indicate a B1/2IIIe spectral type, and all spectral features are at velocities consistent with the Small Magellanic Cloud. At this distance, it is a luminous SSS ($>10^{37}${\lum}) but whose brief peak 
luminosity of $>10^{39}$\lum in the 2--4 keV band makes it the brightest SSS yet seen at ``hard'' X-rays. We propose that \maxi\ is a Be--WD binary, and the first example to possibly enter ULX territory. The brief hard X-ray flash could possibly be a result of the interaction of the ejected nova shell with the B star wind in which the white dwarf (WD) is embedded. This makes \maxi\ only the third Be/WD system in the Magellanic Clouds, but it is by far the most luminous. The properties of \maxi\ give weight to previous suggestions that SSS in nearby galaxies are associated with early-type stellar systems. 

\end{abstract}

\section{Introduction}

Supersoft sources (hereafter SSS) are a class of luminous X-ray sources, so-called because of their very soft ($kT\rm{_{bb}}\lessapprox$ 100~eV) X-ray spectrum, which can reach luminosities up to ${\sim}10^{38}$\lum). Initially discovered in the Magellanic Clouds, as a result of their low interstellar absorption in that direction \citep{long79}, they have subsequently been found in the Milky Way and in nearby galaxies \citep{d2003,d2004,kong2004,kong2005}. 
Key to understanding the nature of the SSS is that their effective blackbody radii are comparable to those of a white dwarf (WD). 
In fact, many hot WDs and pre-WDs have now been observed as SSS including several recent novae, symbiotic systems, and a planetary nebula, in all of which the WD nature of the compact object is well established (see the SSS catalog from \cite{2000NewA....5..137G}). 
This led to \cite{cbss} establishing what is now considered as the SSS paradigm, wherein they are WDs in close binary systems, accreting at very high rates, which leads to (quasi-)stable thermonuclear burning of hydrogen on the WD surface (see also \cite{1994ApJ...426..692R,1996ApJ...470L..97H,1996LNP...472....3D}). 
Furthermore, this requires a massive WD for the most luminous of the class, and so SSS are considered to be strong candidates as progenitors of Type Ia supernovae in the single-degenerate scenario \citep{d2010,compact}. 
SSS are therefore very important for enhancing our understanding of close binary interactions and the late stages of stellar evolution. It also appears that, under rare circumstances, SSS can exhibit ``ultraluminous'' (or ULX) levels, which are difficult to explain through a simple application of the steady nuclear burning model \citep{2008ApJ...674L..73L}. 

While the \textit{RXTE} monitoring programs of the Small Magellanic Cloud (SMC) have revealed an 
extensive population of Be X-ray pulsars (and hence are neutron star (NS) systems, see \cite{2008ApJS..177..189G}), the detection of their WD cousins is hampered by the luminosity of the Be companion and the lack of continuous soft X-ray monitoring of these regions. To date, only two Be+WD systems (the first, XMMU J052016.0-692505, in the LMC \citep{Kahabka06}; the second, XMMU J010147.5-715550, in the SMC \citep{sssinsmc}) have been detected in the Magellanic Clouds. 

In late 2011, \maxi, a new X-ray transient, detectable only at soft ($<$4~keV) energies, was discovered with the MAXI/GSC as a very brief ($<$90 minutes) X-ray flare, in the direction of the Wing of the SMC, but with poor ($\sim0^{\circ}.4$) location accuracy \citep{atel3756}. This X-ray flare (denoted XRF 111111A) exhibited a very unusual spectrum in that all the X-ray flux was confined to the lowest GSC energy channel (2--4~keV), inferring a luminosity at these energies of $>6\times10^{38}$\lum \citep{atel3756} at the distance of the SMC. This was substantially higher than any known Galactic/Magellanic Cloud SSS, and approaching the luminosities seen in some extragalactic SSS \citep{2000NewA....5..137G}.

A \textit{Swift} X-ray and UV/optical observation \citep{atel3758,atel3759} was performed within 10 hr of the MAXI trigger, revealing that \maxi\ was undergoing a supersoft phase of emission ($\sim2\times10^{37}$\lum\ (0.2--2~keV)), and with an accurate ($\pm$3.6 arcsec) location of $\mathrm{R.A.}=01$:59:25.6, $\mathrm{decl.}=-$74:15:28, J2000.0). The X-ray spectra were compatible with a low-temperature blackbody ($kT_\mathrm{bb}\sim$100 eV) that is typical of the SSS. 
\textit{Swift}/UVOT UV/optical images plus archival OGLE-IV \textit{I} band photometry show that the optical counterpart is a bright (13th magnitude in the \textit{U} band) blue star that increased in brightness by at least $\sim$0.5 mag at the time of the MAXI/GSC X-ray flash, and declined on the same timescale as the \textit{Swift}/XRT light curve. 

Ultraluminous X-ray sources (ULXs) are non-nuclear X-ray sources with $L_X{\geq}10^{39}$\lum, whose physical properties are still controversial \citep{2004ApJS..154..519S}. Short-term variability indicates that they must be accreting compact objects, where the mass transfer rate is close to or exceeding the Eddington Limit. Once strong candidates for the long-sought intermediate-mass black holes (IMBHs), they are now considered likely to be extreme examples of stellar-mass black holes (see e.g., \cite{ZampieriRoberts09}). It is thus of considerable current interest as to how accreting WDs can even come close to having properties that appear to overlap with the ULX. 

In this paper, we present a time-resolved, multi-wavelength follow-up of the \maxi\ X-ray flare and its subsequent decline over the next few months, using \textit{Swift XRT/UVOT}, \textit{Galaxy Evolution Explorer} \textit{GALEX}, SAAO 1.9~m, ESO NTT, \textit{Wide-field Infrared Survey Explorer} (\textit{WISE}), and ATCA radio observations to reveal some of the detailed properties of this remarkable object.

\section{Observations}
\subsection{X-Ray: MAXI}
\begin{figure*}[t]
\centering
\includegraphics[width=165mm]{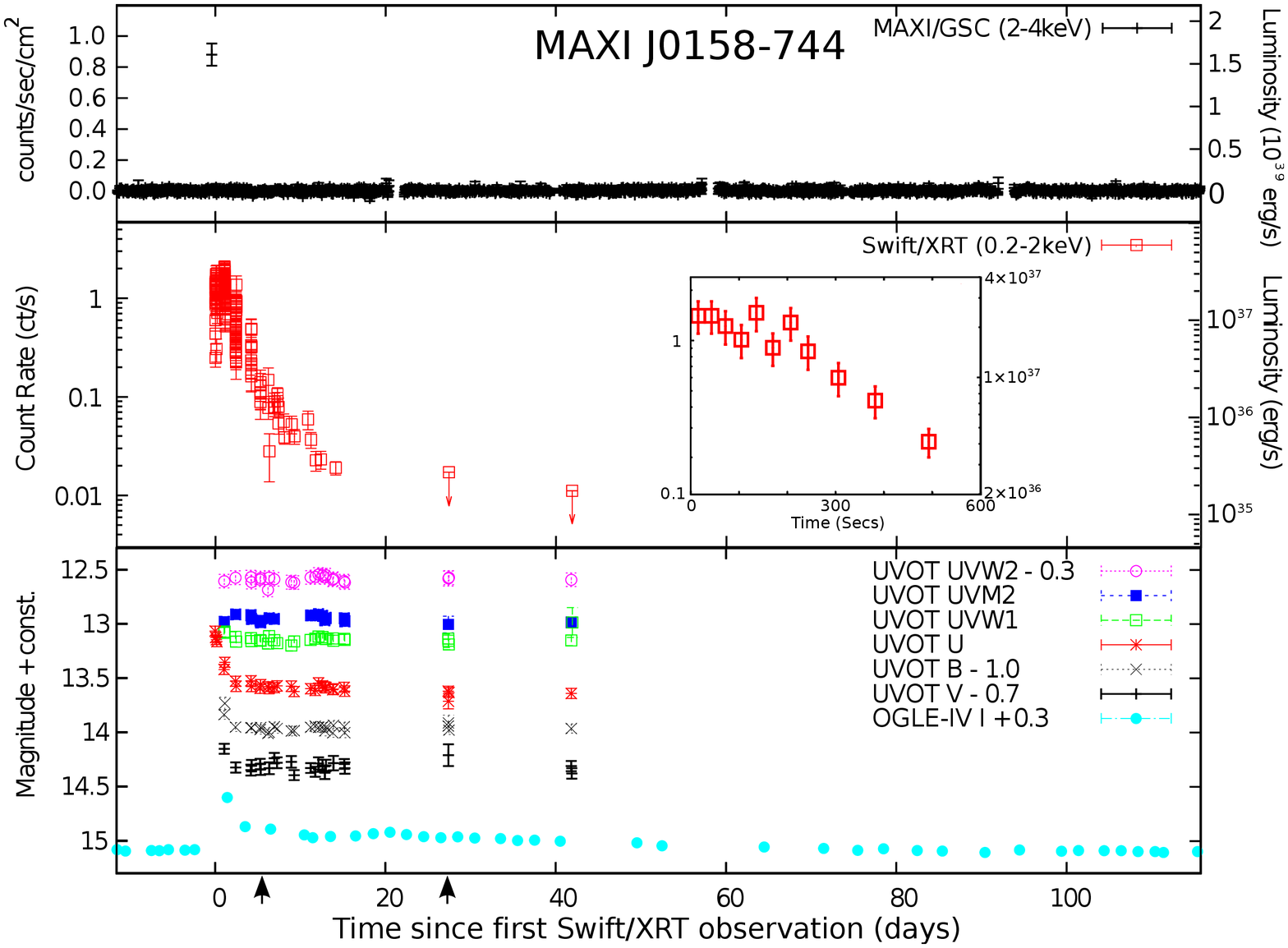}
\includegraphics[width=165mm]{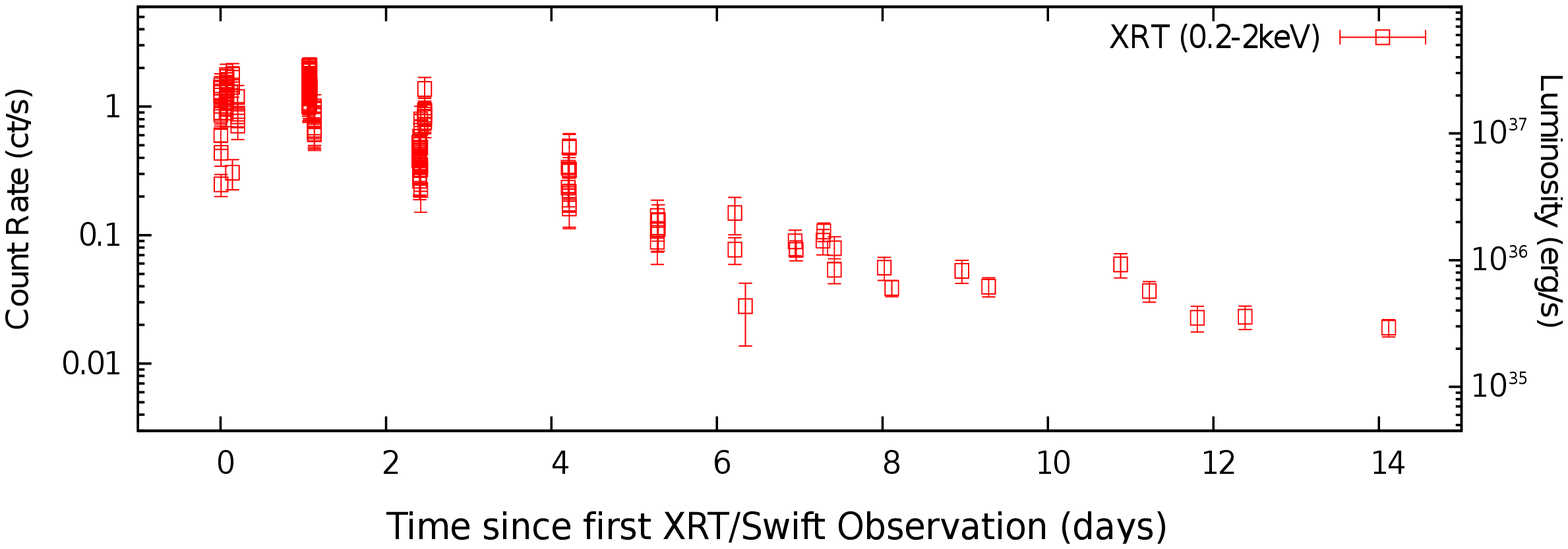}
\caption{
Multi-wavelength lightcurves of \maxi\ from hard X-ray to optical, covering the interval from 2011 October 31 to 2012 March 06, and with the time axis referenced to that of the first \textit{Swift} observation (2011 November 11 15:33 UT). 
First panel: MAXI/GSC 2--4~keV lightcurve with time resolution of 92 minutes (single orbit). The X-ray flash is clearly visible close to time zero, and the X-ray luminosity (right-hand axis) is given for the unabsorbed black-body model and SMC distance (see the text for details); 
Second panel: \textit{Swift}/XRT X-ray lightcurve (0.2--2~keV). The X-ray luminosity (right-hand axis) is given for the X-ray spectral fit corresponding to the peak of the lightcurve and is intended as a guide only (see the text). The inset box shows how the X-ray flux declined by almost an order of magnitude during the first 500~s observation; 
Third panel: \textit{Swift}/UVOT lightcurves covering six bands from the near-UV to optical, plus the OGLE-IV \textit{I}-band lightcurve. 
The two arrows on the abscissa indicate when the optical spectra were taken (2011 November 16 and 2011 December 8). 
Fourth panel: expanded version of the second panel to show the X-ray flux variation more clearly. 
\\}
\label{fig:lc}
\end{figure*}

The MAXI X-ray camera monitors the 0.5--30keV X-ray sky from the International Space Station, providing source alerts and making its data immediately publicly available.\footnote{http://maxi.riken.jp/mxondem/} \maxi\ was detected by the MAXI/GSC on 2011 November 11 as a bright, but short-lived (less than a single 92 minute orbit) soft X-ray flare (for details, see \cite{atel3756}). Using the on-line processed data products provided by the MAXI team (i.e., light curve and spectrum) we show in Figure \ref{fig:lc} (top panel) the MAXI/GSC light curve (at the highest temporal resolution of orbit by orbit) in the lowest energy band provided (2--4~keV). Remarkably, apart from the clear X-ray flare that was detected from 2011 November 11 03:34 to 05:06 UTC with a count rate of $>0.8$ counts~cm$^{-2}$~s$^{-1}$ (Figure \ref{fig:lc}), the source is virtually undetectable at all other times in this 4 month interval. 

The MAXI X-ray spectrum corresponding to the X-ray flare indicates only that it was soft (confined to the lowest energy channel) and so we fitted it with an absorbed blackbody model, and with the column density (\nh) fixed to the galactic value of $4.0\times10^{20}$\cm\ \citep{dlmap}. The best-fit temperature is $380^{+120}_{-80}$~eV, which is consistent with the preliminary result from \cite{atel3756}. 
The low temperature shows that it is a soft X-ray source (but not soft enough to be able to claim that it was in the SSS phase from the outset.) 
Given its location (in the direction of the SMC, and confirmed by our optical spectra below which show it to be at the velocity of the SMC) we assume a distance of 60.6~kpc \citep{2005MNRAS.357..304H}, which leads to an unabsorbed luminosity in the flare of ${\sim}1.6\times10^{39}$\lum\ (2--4~keV). 
(We noted that the difference between the estimated luminosities from this paper (${\sim}1.6\times10^{39}$) and Kimura et al. (2011; 400 mCrab or $6.4\times10^{39}$\lum) is probably due to different estimation methods used, i.e. we estimated the unabsorbed luminosity from X-ray spectral fitting while they estimated it from the source count rate\footnote{http://maxi.riken.jp/top/index.php?cid=000000000036} ). 
This immediately established great interest in \maxi\ as potentially the most luminous SSS in the Magellanic Clouds ever reported (see \cite{2000NewA....5..137G} for a compilation of known SSS). Furthermore, this extremely high luminosity places \maxi\ into the ULX regime, bearing comparison with sources usually interpreted as high-mass transfer black hole (BH) systems (see \cite{2012ApJ...747L..39H,2012ApJ...750..152S}). 

For comparison with the blackbody model, we also fitted a Raymond--Smith thermal plasma model. This gave a similar unabsorbed luminosity, and with a best-fit temperature of ${\sim}0.9$~keV.
The signal-to-noise ratio (S/N) of the MAXI spectrum does not allow a distinction between the models (there are only six pulse-height analyzer bins within the 2--4~keV range).

\subsection{X-Ray: \textit{Swift}/XRT}
\begin{table}[h]
\centering
\caption{\textit{Swift} Observations of \maxi}
\scriptsize
\begin{tabular}{ccccc}
\hline
Obs Id & Date\footnote{All observations were taken in 2011. } & Exp(XRT) & Exp(UVOT) & Mode\footnote{PC: photon-counting; WT: window-timing. } \\ 
& (m/d) (MJD) & (s) & (s) & (PC/WT) \\ 
\hline
32181001 & Nov 11 (55876) & 682 & 867 & PC\\
32182001 & Nov 11 (55876) & 735 & 912 & PC\\
32187001 & Nov 12 (55877) & 1994 & 1950 & WT\\
32187002 & Nov 14 (55879) & 1994 & 1956 & WT\\
32187004 & Nov 15 (55880) & 2024 & 1966 & WT\\
32187005 & Nov 16 (55881) & 2169 & 2111 & WT\\
32187006 & Nov 17 (55882) & 1724 & 1676 & WT\\
32187007 & Nov 18 (55883) & 2146 & 2097 & PC\\
32190001 & Nov 19 (55884) & 4351 & 4316 & PC\\
32190002 & Nov 19 (55884) & 707 & 667 & PC\\
32187008 & Nov 20 (55885) & 2567 & 2517 & PC\\
32187010 & Nov 22 (55887) & 2292 & 1570 & PC\\
32187011 & Nov 23 (55888) & 2337 & 2286 & PC\\
32187012 & Nov 24 (55889) & 2339 & 2262 & PC\\
32187013 & Nov 25 (55890) & 1354 & 1316 & PC\\
32187014 & Nov 26 (55891) & 2448 & 2398 & PC\\
32187015 & Dec 09 (55904) & 1983 & 1925 & PC\\
32187016 & Dec 23 (55918) & 3677 & 2481 & PC\\
\hline
\end{tabular}
\label{tab:swift}
\end{table}


\maxi\ was observed by \textit{Swift} between 2011 November and December (see Table \ref{tab:swift}). 
The first two observations were taken in photon counting (PC) mode in order to accurately locate the source, since the MAXI all-sky camera only provides 0$^{\circ}$.42 plus 0$^{\circ}$.1 (systematic uncertainty) positional information \citep{atel3756}. Subsequent observations were taken in window timing (WT) mode, in order to reduce pile-up effects due to the high initial count rate ($>$0.6 counts s$^{-1}$), and then returned to PC mode for the last 11 observations. Data calibration, extraction, and event filtering were performed according to standard procedures outlined in the \textit{Swift}/XRT data reduction guide \citep{capalbi}. 


In contrast with the behavior of the well-known class of Galactic soft X-ray transients/X-ray novae (which are all BH/NS systems, see \cite{2006csxs.book..215C}), the outburst of this source lasted barely two weeks. Within a month it was below the \textit{Swift} detection limit, indicating a two orders of magnitude drop in intensity (Figure \ref{fig:lc}). In order to search for any periodicities in the data, we performed a Fourier analysis of the MAXI and \textit{Swift} data on all relevant timescales. However, no significant signal was present in the power spectra. 

We then established different time intervals for analysis based on the flux levels in the overall light curve, producing summed spectra for each time period of interest.
As the first two observations in Table \ref{tab:swift} are essentially a single observation, we merged them to form a combined spectrum. 
In the following three observations (November 12--15), the X-ray source was still strong, and therefore we grouped them individually. 
We combined the remaining observations as follows in order to increase the S/N: November 16--17; November 18--20,22--26. 
The final two observations were not used, as the source, while marginally detected, was too weak to produce useable spectra. 

\subsubsection{XRT Spectral Fitting}
\begin{figure*}[t]
\centering
\includegraphics[width=165mm]{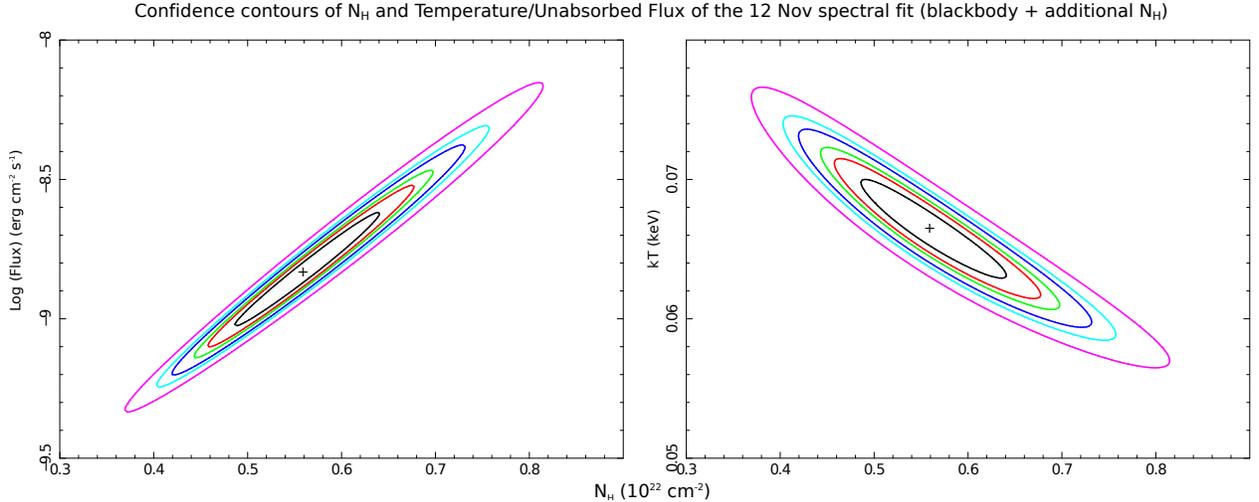}
\caption{
$\chi^2$ contours in the (\nh, unabsorbed flux) plane (left) and in the (\nh, $kT$) plane (right) of the November 12 spectral fit (blackbody + additional \nh). The global minimum (or best-fit) point marked as a black cross near the center ($\chi^2$ = 60 for 55 dof, see Table \ref{tab:spec}). The contours are plotted at $\Delta \chi^2$ = 2.30, 4.61, 6.17, 9.21, 11.8, and 18.4, equivalent to the 1$\sigma$ , 90\%, 2$\sigma$ , 99\%, 3$\sigma$ , and 99.99\% confidence levels for two parameters. \\}
\label{fig:nhcontour}
\end{figure*}

\begin{figure*}[t]
\centering
\includegraphics[width=165mm]{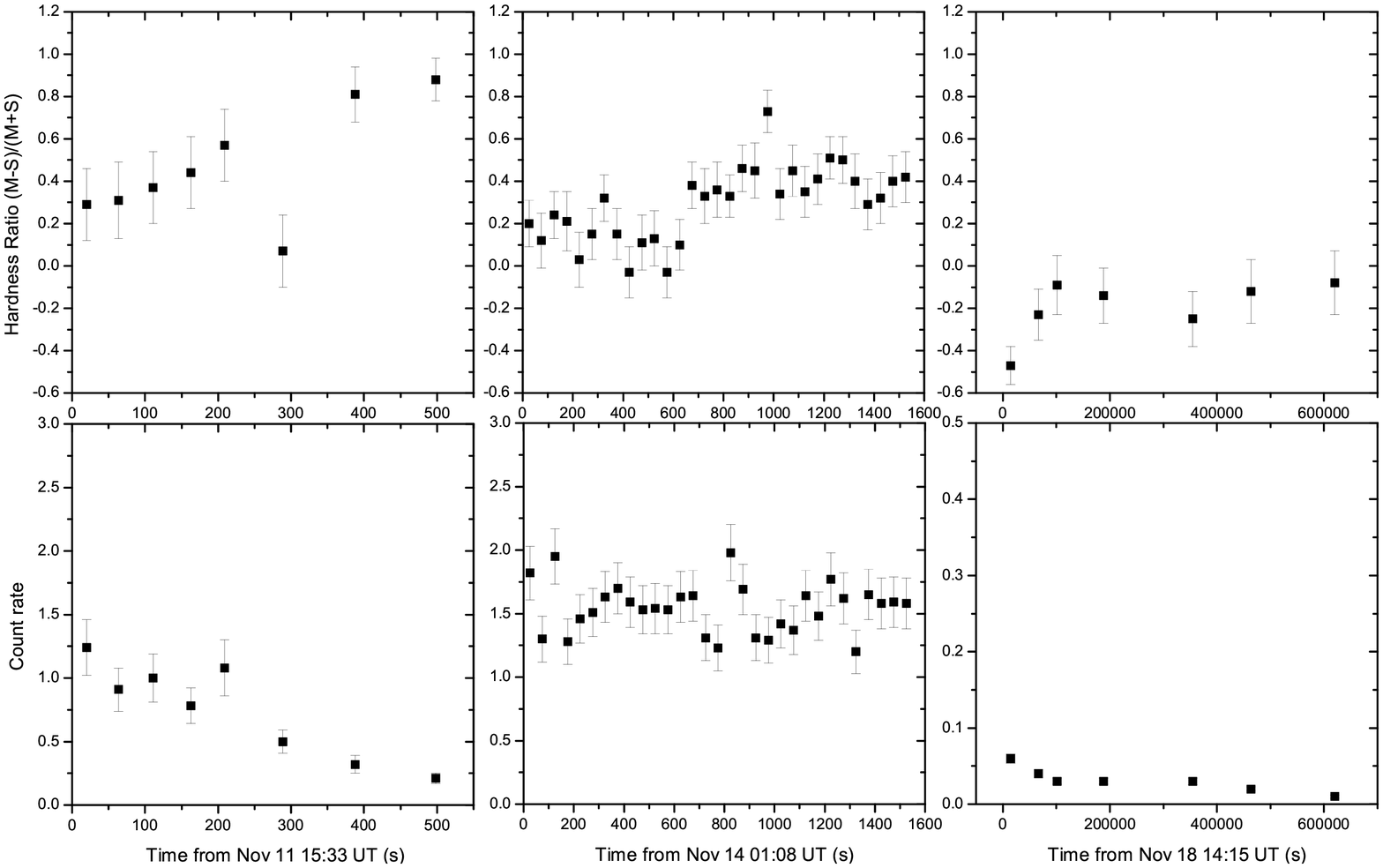}
\caption{Demonstration of the X-ray spectral variability exhibited by \maxi. Upper panels: plot of hardness ratio, defined as (M$-$S)/(M+S), against time; lower panels: corresponding X-ray light curve. The first and the third boxes show a tentative correlation between the X-ray intensity and hardness ratio on quite different timescales, while the middle box shows a clear hardness ratio jump, which occurs when the 0.7~keV emission feature turns on (see the text). \\}
\label{fig:hr}
\end{figure*}

The X-ray spectra were fitted with \texttt{XSPEC} \citep{1996ASPC..101...17A} using mostly blackbody models, but we also tried the T\"{u}bingen WD model atmospheres \citep{2003ASPC..288..103R,1999JCoAM.109...65W,2003ASPC..288...31W} which can be used in \texttt{XSPEC} as a tabular grid of models for fitting. Nevertheless, all the temperatures relevant here are low ($\sim$100eV), which makes the results (especially the implied luminosity) extremely sensitive to the \nh\ values of the fit, if the latter become high. 
This is demonstrated in Figure \ref{fig:nhcontour}, which shows $\chi^2$ contour maps of the November 12 spectral fit (blackbody + additional \nh), indicating how the (\nh, unabsorbed flux) and (\nh, $kT$) correlate with each other, leading to potentially large systematic errors in the unabsorbed flux. This is clear in the results summarized in Table \ref{tab:spec}.
Accordingly, we assumed the presence of a Galactic \nh\ value of $4\times10^{20}$\cm \citep{dlmap} in all models, so as to avoid overestimating the unabsorbed flux or underestimating the temperature. 
All the XRT spectra give temperatures in the range 80--120~eV (see top half of Table \ref{tab:spec}). The inferred luminosities decrease exponentially from the initial peak, and the source drops below the \textit{Swift} detection limit about a month later (Figure \ref{fig:lc}). However, it should be noted that the source and its spectrum varied dramatically from early in the outburst (as demonstrated by the hardness ratio (HR) and X-ray flux plots (Figure \ref{fig:hr} and the inset in Fig \ref{fig:lc}). Hence a blackbody may not be appropriate to account for all phases of the outburst. Higher resolution observations (e.g., \textit{CXO} \citep{2000SPIE.4012....2W}) do indicate the importance of using more physically relevant models, even though \textit{Swift}/XRT's limited sensitivity, spectral and temporal resolution cannot directly distinguish among them. 
Nonetheless, the fit parameters are still useful for comparison with the behavior of other SSS. 

In addition to the fixed \nh\ blackbody fitting, we tried adding an additional absorbing component to see if there is any intrinsic absorption in the system. 
This extra absorption had abundances set to those appropriate for the SMC 
(i.e. He=0.83; C=0.13, N=0.05; O=0.15; Ne=0.19; Na=0.41; Mg=0.24; Al=0.79 Si=0.28; S=0.21; Cl=0.28; Ar=0.16; Ca=0.21; Cr=0.26; Fe=0.20; Co=0.20; and Ni=0.40, relative to solar) \citep{smcabun}. 
Except for the last spectrum (that required no additional absorption), the fits to all other spectra were significantly improved. 
However, many of the parameters are not then well constrained, especially the normalization factors (i.e., $\mathrm{norm} = R_{\mathrm{km}}^2/D_{10}^2$ where $R_{\mathrm{km}}$ is the source radius in km, and, $D_{10}$ is the distance to the source in units of 10 kpc\footnote{https://heasarc.gsfc.nasa.gov/xanadu/xspec/manual/XSmodelBbodyrad.html}), and so these are given as upper limits in Table \ref{tab:spec}. 

\begin{figure*}[t]
\centering
\fbox{
\includegraphics[width=210mm,angle=90]{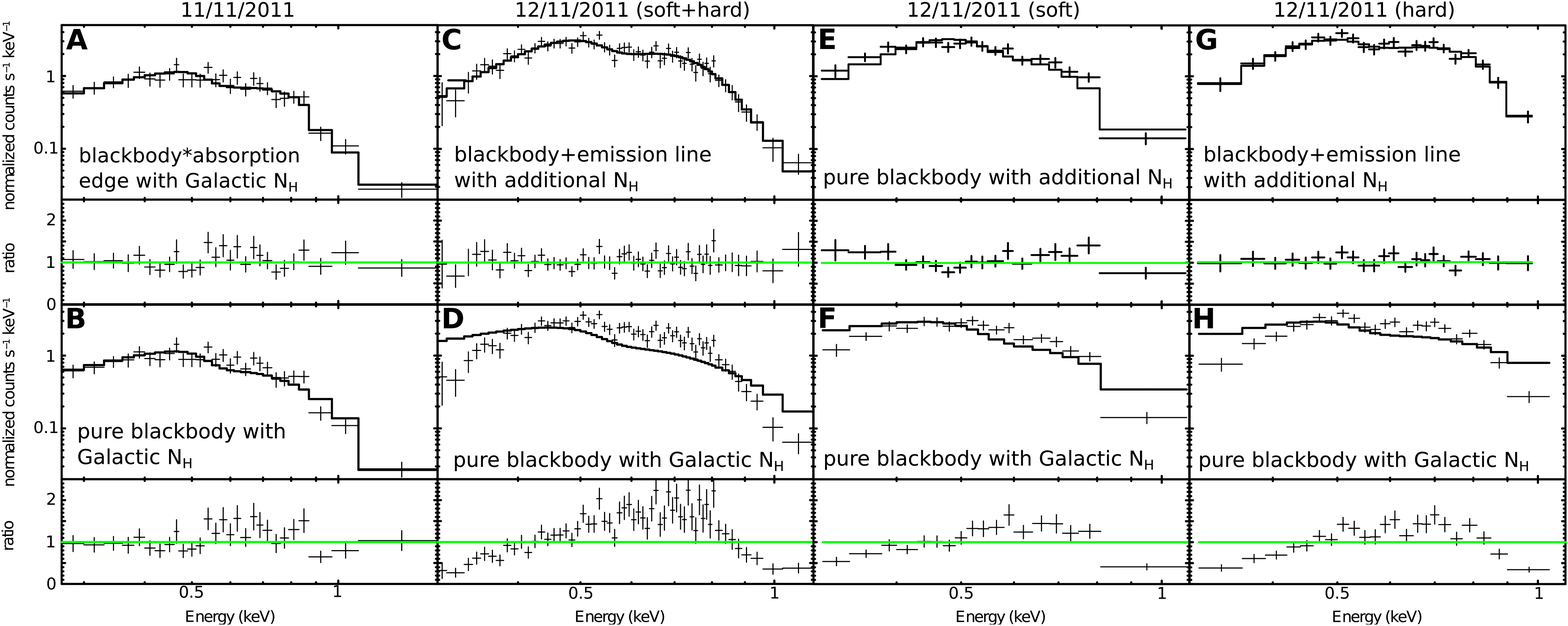}
}
\fbox{
\includegraphics[width=210mm,angle=90]{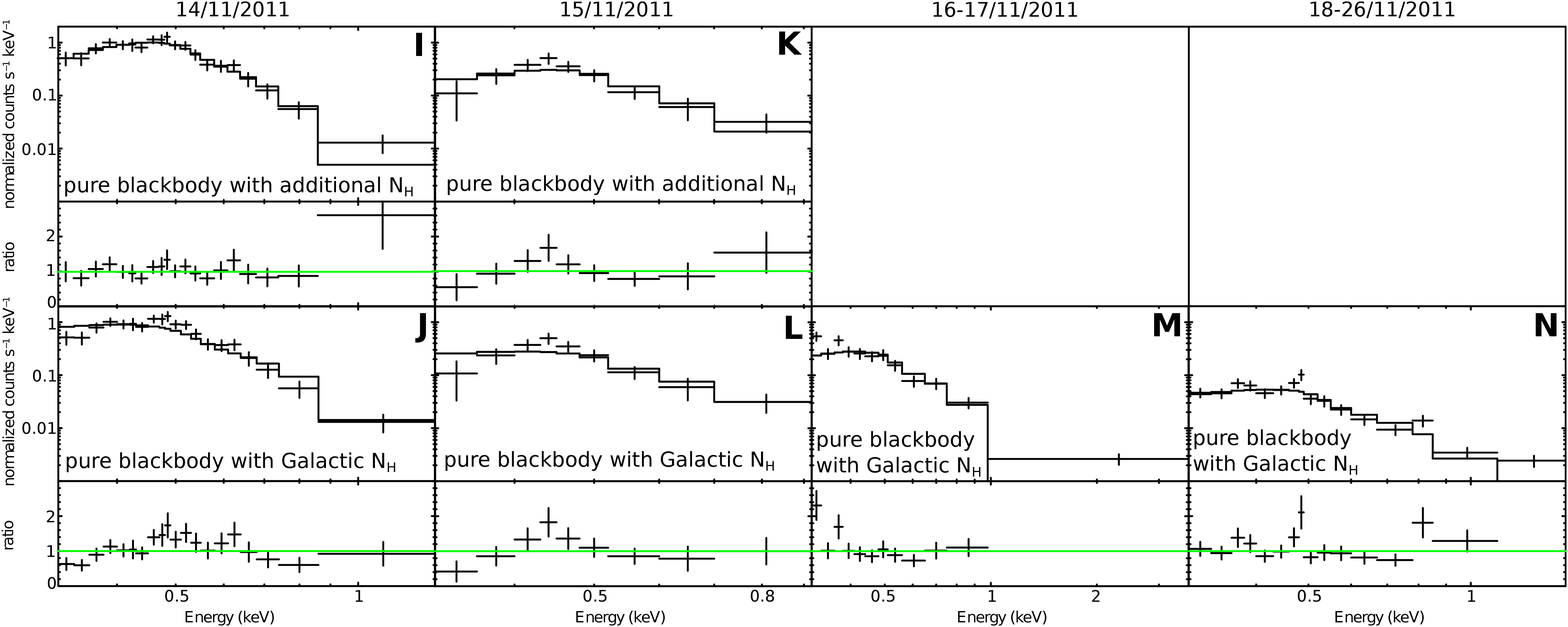}
}
\caption{
Upper panel: early outburst \textit{Swift}/XRT spectra of \maxi\ with their best-fit models (below which are plotted the residuals to the fits): 
\\Left column: (A) November 11 spectrum with blackbody + absorption edge at 0.89keV (and fixed Galactic \nh), and (B) pure blackbody with fixed Galactic \nh\ of $4\times10^{20}$\cm; 
\\Second column: (C) November 12 spectrum with blackbody plus an emission line with additional \nh, and (D) pure blackbody with fixed Galactic \nh\ of $4\times10^{20}$\cm. 
\\Right two columns show the spectra and fits when dividing the November 12 spectrum to be before (soft) and after (hard) the HR jump seen in Figure \ref{fig:hr}. The ``hard'' spectrum (G and H) includes a broad emission feature at 0.7keV. Note that the absorption and emission features are visible in the residuals to the simple fits. 
\\Lower panel: late outburst \textit{Swift}/XRT spectra (after November 12) of \maxi\ with their best-fit blackbody models (with or without additional \nh). No valid fits have been obtained by using a blackbody with additional \nh\ for the November 16--17 and November 18--26 Nov spectra. 
\\}
\label{fig:spec}
\end{figure*}

We also attempted to find a better model by trying various model combinations.
We found that an absorption edge at 0.9~keV and an emission line at 0.7~keV\footnote{It has been suggested that the 0.7~keV excess could be explained by an overabundance of oxygen and neon along the line of sight \citep{neon}. } can ameliorate the fits of November 11 and 12, respectively. 
To estimate the significances of these spectral features, we performed \textit{F}-tests\footnote{See the Appendix of \cite{2012ApJ...748...86O} for details. } on the fits, which give about 85\% confidence for the edge and $>$99\% confidence (close to $3\sigma$) for the emission line. 
The best fit parameters for these models are shown in parentheses in Table \ref{tab:spec}, and all the models plotted in Figure \ref{fig:spec}. 

In addition to the detailed spectral analysis, we investigated both short and long timescale spectral variability in \maxi\ by plotting the spectral HR (defined as $\mathrm{HR=(M-S)/(M+S)}$) against the X-ray lightcurve. The soft (S) and medium bands (M) cover 0.2--0.5 keV and 0.5--2.0 keV, respectively. We find (Figure \ref{fig:hr}) that HR varied throughout most of the observations, especially in the early and late times of the outburst, when the changes are correlated with the X-ray flux (i.e., a lower count rate gives a harder spectrum). 
Furthermore, we noted a jump in HR on 2011 November 12 at 17:16 UTC from $0.13\pm0.05$ to $0.41\pm0.10$ which is clearly visible in the middle panel of Figure \ref{fig:hr}. 
We divided the spectra accordingly and found that the jump was associated with the sudden appearance of the 0.7~keV emission feature required by the fits (Figure \ref{fig:spec}). 

\begin{table*}
\centering
\scriptsize
\caption{ \textit{Swift}/XRT Spectral Properties of \maxi}
\begin{tabular}{@{}lllllcl}
\hline
Date & \nh\ (Additional) & Unabsorbed & Blackbody & Radius & & Notes\\ 
& & $L_X$(0.2--2keV)\footnotemark[1] & Temperature & & & \\ 
& ($10^{20}$\cm) & ($10^{37}$\lum) & ($\mathrm{eV}$) & ($\mathrm{10^{9} cm}$) & $\chi^2$/dof & \\ 
\hline
\multicolumn{7}{c}{Blackbody + fixed \nh\ of $4{\times}10^{20}$\cm (\texttt{wabs*bbodyrad})} \\
\hline
2011 Nov 11 \T \B & \nodata & $2.00^{+0.12}_{-0.14}$ & $108^{+3}_{-3}$ & $0.12^{+0.01}_{-0.01}$ & 40/26 & \nodata \\ 
2011 Nov 12 & \nodata & $2.31^{+0.07}_{-0.10}$ & $114_{-3}^{+3}$ & $0.11^{+0.01}_{-0.01}$ & 457/56 & \nodata \\ 
2011 Nov 12 (soft) & \nodata & $2.51^{+0.11}_{-0.21}$ & $101^{+4}_{-3}$ & $0.15^{+0.02}_{-0.01}$ & 184/34 & \nodata \\
2011 Nov 12 (hard) & \nodata & $2.36^{+0.10}_{-0.16}$ & $130^{+5}_{-4}$ & $0.08^{+0.01}_{-0.01}$ & 306/44 & \nodata\\
2011 Nov 14 & \nodata & $0.98^{+0.06}_{-0.18}$ & $81^{+4}_{-4}$ & $0.16^{+0.03}_{-0.02}$ & 40/18 & \nodata \\ 
2011 Nov 15 & \nodata & $0.48^{+0.06}_{-0.34}$ & $83^{+11}_{-10}$ & $0.10^{+0.05}_{-0.03}$ & 12/7 & \nodata \\ 
2011 Nov 16--17 & \nodata & $0.27^{+0.02}_{-0.04}$ & $95^{+6}_{-5}$ & $0.06^{+0.01}_{-0.01}$ & 39/10 & \nodata \\ 
2011 Nov 18--26 & \nodata & $0.09^{+0.01}_{-0.01}$ & $92^{+6}_{-5}$ & $0.03^{+0.01}_{-0.01}$ & 44/14 & \nodata \\ 
\hline
\multicolumn{7}{c}{Blackbody + additional \nh\ (\texttt{phabs*vphabs*bbodyrad})\footnotemark[4]} \\
\hline
2011 Nov 11\footnotemark[5] \T \B & $5^{+6}_{-4}$ ($\lesssim5$) & $2.71^{+0.15}_{-0.61}$ ($1.90^{+0.04}_{-0.76}$) & $102^{+7}_{-7}$ ($122^{+4}_{-11}$) & $0.15^{+0.06}_{-0.04}$ ($0.09^{+0.04}_{-0.01}$) & 38/25 (23/23) & (+edge at $0.89^{+0.03}_{-0.03}$~keV) \\ 
2011 Nov 12\footnotemark[6] & $56^{+9}_{-8}$ ($48^{+11}_{-11}$) & $65^{+1}_{-41}$ ($45.0^{+0.2}_{-39.7}$) & $66^{+4}_{-4}$ ($67^{+5}_{-6}$) & $2.1^{+1.1}_{-0.7}$ ($\lesssim1.7$) & 62/55 (47/52) & (+line at $0.73^{+0.06}_{-0.11}$~keV) \\ 
2011 Nov 12 (soft) & $44^{+13}_{-10}$ & $\lesssim45.2$ & $64^{+6}_{-5}$ & $\lesssim3.6$ & 50/33 & \nodata \\
2011 Nov 12 (hard)\footnotemark[6] & $64^{+13}_{-11}$ ($54^{+22}_{-14}$) & $\lesssim83.1$ ($55^{+1}_{-50}$) & $68^{+5}_{-5}$ ($67^{+9}_{-13}$) & $\lesssim2.2$ ($1.9^{+5.9}_{-1.0}$) & 49/43 (34/40) & (+line at 0.7~keV (fixed)) \\
2011 Nov 14 & $30^{+17}_{-12}$ & $\lesssim11.8$ & $56^{+9}_{-9}$ & $\lesssim1.4$ & 13/17 & \nodata \\ 
2011 Nov 15 & $19^{+37}_{-17}$ & $\lesssim2.7$ & $62^{+21}_{-28}$ & $\lesssim0.5$ & 9/6 & \nodata \\ 
2011 Nov 16--17\footnotemark[7] & \nodata & \nodata & \nodata & \nodata & \nodata & \nodata \\ 
2011 Nov 18--26\footnotemark[7] & \nodata & \nodata & \nodata & \nodata & \nodata & \nodata \\ 
\hline
\multicolumn{7}{c}{WD atmosphere model + fixed \nh\ of $4{\times}10^{20}$\cm (\texttt{wabs*tmap})\footnotemark[8]} \\
\hline
2011 Nov 11 \T \B & \nodata & $1.55^{+0.07}_{-0.33}$ & $90.5$ (max) & \nodata & 212/26 & \nodata \\ 
2011 Nov 12 & \nodata &$1.87^{+0.03}_{-0.25}$ & $90.5$ (max) & \nodata & 698/56 & \nodata \\ 
2011 Nov 12 (soft) & \nodata & $2.11^{+0.05}_{-0.31}$ & $90.5$ (max) & \nodata & 194/34 & \nodata \\
2011 Nov 12 (hard) & \nodata & $1.73^{+0.05}_{-0.37}$ & $90.5$ (max) & \nodata &557/44 & \nodata\\
2011 Nov 14 & \nodata & $0.70^{+0.03}_{-0.10}$ & $90.5$ (max) & \nodata & 21/18 & \nodata \\ 
2011 Nov 15 & \nodata & $0.34^{+0.03}_{-0.12}$ &$90.5$ (max) & \nodata & 12/7 & \nodata \\ 
2011 Nov 16--17 & \nodata & $0.21^{+0.01}_{-0.05}$ & $90.5$ (max) & \nodata & 57/10 & \nodata \\ 
2011 Nov 18--26 & \nodata & $0.06^{+0.01}_{-0.01}$ & $90.5$ (max) & \nodata & 63/14 & \nodata \\ 
\hline
\multicolumn{7}{c}{WD atmosphere model + additional \nh\ (\texttt{phabs*vphabs*tmap})\footnotemark[8]} \\
\hline
2011 Nov 11 \T \B & $31^{+5}_{-5}$ & $9.6^{+0.2}_{-2.5}$ & $90.5$ (max) & \nodata & 173/25 & \nodata \\ 
2011 Nov 12 & $51^{+4}_{-3}$ & $20.9^{+0.1}_{-4.4}$ & $90.5$ (max) & \nodata & 139/55 & \nodata \\ 
2011 Nov 12 (soft) & $29^{+5}_{-5}$ & $8.6^{+0.1}_{-2.2}$ & $90.5$ (max) & \nodata & 62/33 & \nodata \\
2011 Nov 12 (hard) & $67^{+5}_{-5}$ & $37.5^{+0.2}_{-14.7}$ & $90.5$ (max) & \nodata & 109/43 & \nodata\\
2011 Nov 14 & $\lesssim11$ & $0.90^{+0.10}_{-0.11}$ & $90.5$ (max) & \nodata & 18/17 & \nodata \\ 
2011 Nov 15\footnotemark[7] & \nodata & \nodata & \nodata & \nodata & \nodata & \nodata \\ 
2011 Nov 16--17\footnotemark[7] & \nodata & \nodata & \nodata & \nodata & \nodata & \nodata \\ 
2011 Nov 18--26\footnotemark[7] & \nodata & \nodata & \nodata & \nodata & \nodata & \nodata \\ 
\hline
\end{tabular}
\footnotetext{Uncertainties of the unabsorbed luminosities are estimated by approximating $\Delta\,L_\mathrm{unabs}$/$L_\mathrm{unabs}\approx\Delta\,L_\mathrm{obs}$/$L_\mathrm{obs}$. }
\footnotetext{Upper limits are shown for those parameters with lower limits pegged to zero OR where irrational errors were obtained (i.e. best-fit value lies outside the error range) during 90\% confidence interval calculations. }
\footnotetext{For those fits with $\chi^2_{\nu}>2$, we used \texttt{cstat} to re-fit and calculate the errors. }
\footnotetext{\nh\ of $4\times10^{20}$\cm\ with XSPEC's built-in solar abundances \citep{1982GeCoA..46.2363A} was included as a fixed contribution to the absorption within the Galaxy. }
\footnotetext{An absorption edge was included in the model where the best-fit threshold energy is shown in the \texttt{Notes} column and the resulting parameters are shown in parentheses. }
\footnotetext{An additional Gaussian emission line is included in the model where the best-fit emission energy is shown in the \texttt{Notes} column and the resulting parameters are shown in parentheses. }
\footnotetext{No valid fit was found. }
\footnotetext{All best-fit temperatures reach the maximum value (1.05 million K) allowed for the WD atmosphere model. }

\label{tab:spec}
\end{table*}

\subsection{UV/Optical: \textit{Swift}/UVOT}
Simultaneously with the\textit{ Swift/XRT} observations, UVOT images in six UV to optical filters were obtained, with $\sim100$~s per exposure (except for the very first observation, that was \textit{U}-band only). We performed aperture photometry on these data, using standard UVOT procedures \citep{2005SSRv..120...95R} to obtain lightcurves. 
To optimize the S/N, we used an aperture radius of 3\arcsec as recommended by HEASARC \citep{2005SSRv..120..165B} and a source-free background region that minimizes contamination from nearby sources. 

We also obtained a single 5000~s UVOT optical spectrum (2800--5200~\AA{}) on 2011 November 19. This was processed using standard UVOT procedures and background subtracted. Wavelength calibration is limited by the extended shape of the zeroth-order image to an accuracy $\sim10$~\AA{}, and a systematic offset of up to 66~\AA{} (see \cite{uvot} for details). There are no obvious emission/absorption features present, but it is brighter at shorter wavelengths, as indicated by the photometry.

\subsubsection{UVOT Results}
In the early part of the \maxi\ outburst, our \textit{UVW1}, \textit{U}, \textit{B}, and \textit{V} images show significant brightening, by up to $\sim$0.5 mag, followed by a rapid decline (within days) to a stable level (Figure \ref{fig:lc}). 
In contrast, the \textit{UVM2} and \textit{UVW2} levels stayed almost constant, at $UM2\sim13.0$ and $UW2\sim12.9$ (with uncertainty $\sim$0.05 mag during the entire \textit{Swift} campaign). 
Assuming the Galactic column of \nh$\,=4.0\times10^{20}$\cm, we estimated the \textit{V}-band extinction to \maxi\ as $A_V=0.18$ using \nh/$A_V$=$2.21\times10^{21}$\cm$\,$mag$^{-1}$ \citep{2009MNRAS.400.2050G}. 
We then used the extinction curve in \cite{uvextinction} to estimate the extinction values for the other bands: $A_B=0.24$, $A_U=0.29$, $A_{UW1}=0.39$, $A_{UM2}=0.55$, and $A_{UW2}=0.48$. 
(Some data were excluded from this analysis as a result of spacecraft pointing drifts, which clearly produced spurious results.)

\subsection{Optical: OGLE-IV Monitoring}

\maxi\ is located in a region of the SMC that is monitored by the OGLE-IV project \citep{2008AcA....58..187U}. Consequently, this provides us with close to nightly monitoring of the source, both before and during the interval covered by the \textit{Swift} observations. As expected, there is a bright optical source present in the Optical Gravitational Lensing Experiment (OGLE) images, at $\mathrm{R.A.}=01$:59:25.87, $\mathrm{decl.}=-$74:15:28.0 (J2000.0), which is well within the \textit{Swift}/XRT error circle. 
The OGLE data are restricted to the \textit{I}-band, and are plotted as the bottom curve of Figure \ref{fig:lc}. A major \textit{I}-band flare (reaching $I=14.3$) is clearly seen at the time of the X-ray flash. In fact, this point was obtained 1--2 days after the X-ray flash, which means that the peak I-band magnitude could be even higher. 

From the OGLE monitoring prior to the data plotted here, we find that $I=14.82\pm0.01$ from 500 days to 320 days before the X-ray flash. At this point, the optical light increased to a new quasi-quiescent magnitude ($14.79\pm0.01$), which persisted up to the major flare. 

Besides the major flare, we also note a secondary peak almost exactly 20 days after the main outburst. Curiously, there is a small-scale ``blip'' in the MAXI/GSC lightcurve at the same time (Figure \ref{fig:lc}). However, the S/N of the MAXI/GSC data here is so weak ($\mathrm{S/N}=1.7$), that we cannot claim this is a real signal. 
Unfortunately, no \textit{Swift XRT/UVOT} observations were carried out during that period. 
The \textit{I}-band flux finally returned to the pre-outburst level $\sim$60 days after the X-ray flash. 

\subsection{Optical Spectroscopy: SAAO/1.9m}

\begin{figure*}
\centering
 \includegraphics[width=180mm]{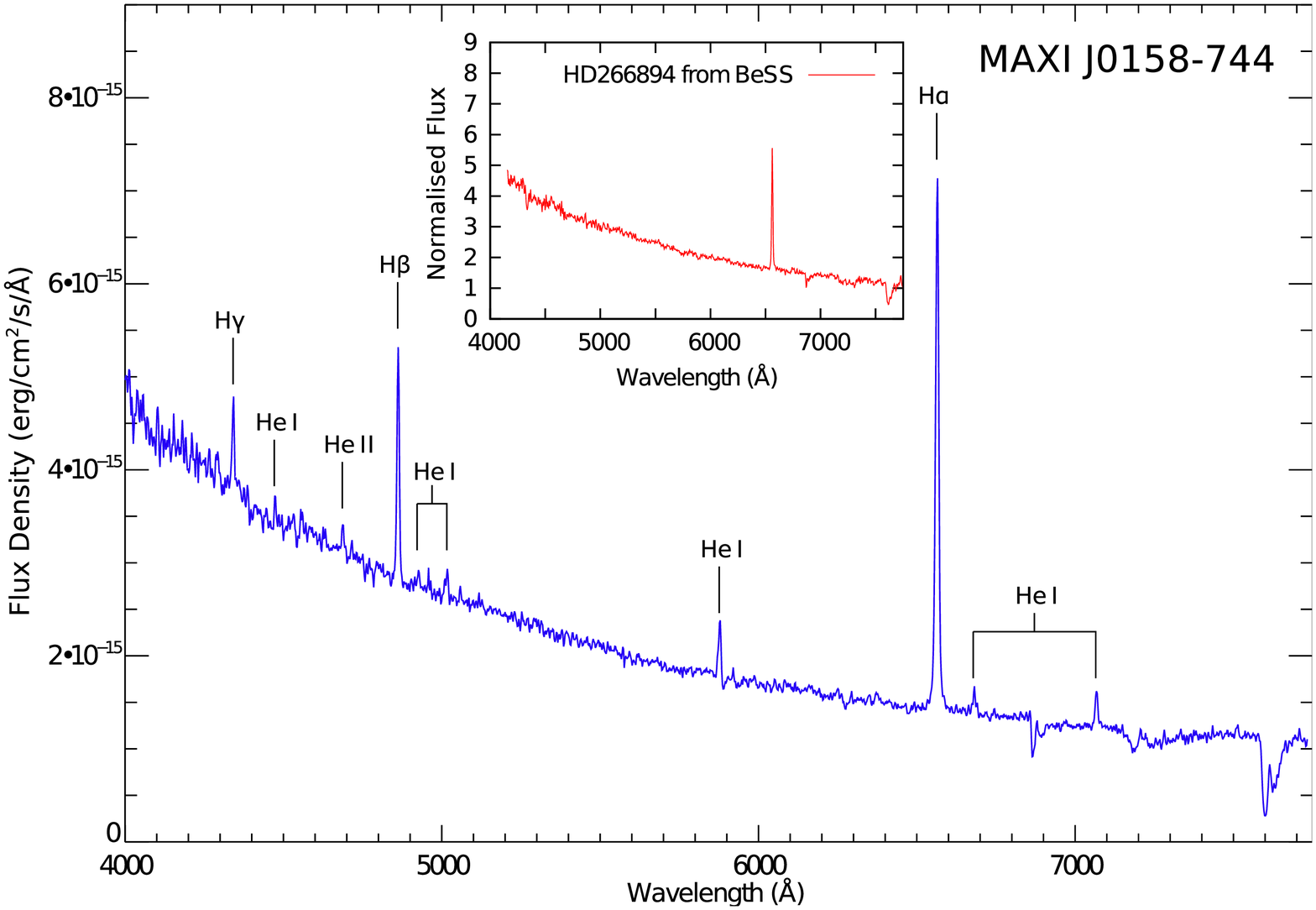}
\caption{Low-resolution optical spectrum of \maxi\ obtained with the SAAO/1.9m on 2011 November 16. The spectrum has been flux calibrated and redshift corrected by $-158$~km~s$^{-1}$. Dominant emission features are marked. The inset shows a Be star (HD 266894) spectrum from BeSS for comparison. \\}
\label{fig:optical_wide}
\end{figure*}

A low resolution, wide wavelength ($\lambda\lambda$4000--7700) optical spectrum of \maxi\ was obtained with the SAAO 1.9~m telescope and CCD spectrograph at Sutherland, South Africa on 2011 November 16, beginning at 22:23UT, approximately 5 days after the initial X-ray flash. Two 1800s exposures using the SITe CCD camera and 300 $\mathrm{l\,mm^{-1}}$ grating (giving 7\AA\ resolution) were obtained within a timespan of 2 hours, and the average of these is plotted in Figure \ref{fig:optical_wide}. It is dominated by very strong Balmer emission, but also exhibits strong \hei\ emission at $\lambda$5876 and $\lambda$6678, and weaker \heii$\lambda$4686 emission. 
\heii$\lambda$4686 is usually present in SSS systems (e.g., CAL83; \citealt{1987ApJ...321..745C} and V407 Cyg; \citealt{2012ApJ...748...43N}), therefore such emission provides further confirmation of the identification of this object with the X-ray outburst. While the spectrum has characteristics of SSS, it does contrast with that of the prototype SSS, CAL83, where \heii$\lambda$4686 is by far the dominant emission feature in the spectrum \citep{1987ApJ...321..745C}.

\subsection{Optical Spectroscopy: ESO/FOS}

Optical spectra were taken on 2011 December 8 (almost 1 month after the initial X-ray flash) with the ESO Faint Object Spectrograph (EFOSC2) mounted at the Nasmyth B focus of the 3.6m NTT. A slit width of 1\arcsec.5 was employed, together with a 600~lines~mm$^{-1}$ grating that yielded 1~\AA{}~pixel$^{-1}$ dispersion over a wavelength range of $\lambda\lambda3095$--$5085$~\AA. The resulting spectra were recorded on a Loral/Lesser, thinned, AR-coated, UV-flooded, MPP CCD with 2048$\times$2048 pixels, at a spectral resolution of $\sim10$~\AA{}. The data were reduced using the standard packages available in the Image Reduction and Analysis Facility (IRAF). Wavelength calibration was achieved with helium and argon arc lamps, and the data were reduced using IRAF. The resulting spectrum of MAX J0158-744 was normalized to remove the continuum and then redshift corrected for an assumed SMC recession velocity of $-$158~km~s$^{-1}$.

\subsubsection{Spectral Classification}

OB stars in the Milky Way are classified using certain metal and helium line ratios \citep{Walborn90}, but this cannot be applied in lower metallicity environments, such as the SMC, due to the weakness or the absence of the metal lines. Accordingly, we classified the spectrum of MAXI J0158-744 using the method of \citet{Lennon97} for B-type stars in the SMC.

\begin{figure*}
\centering
 \includegraphics[height=180mm,angle=90]{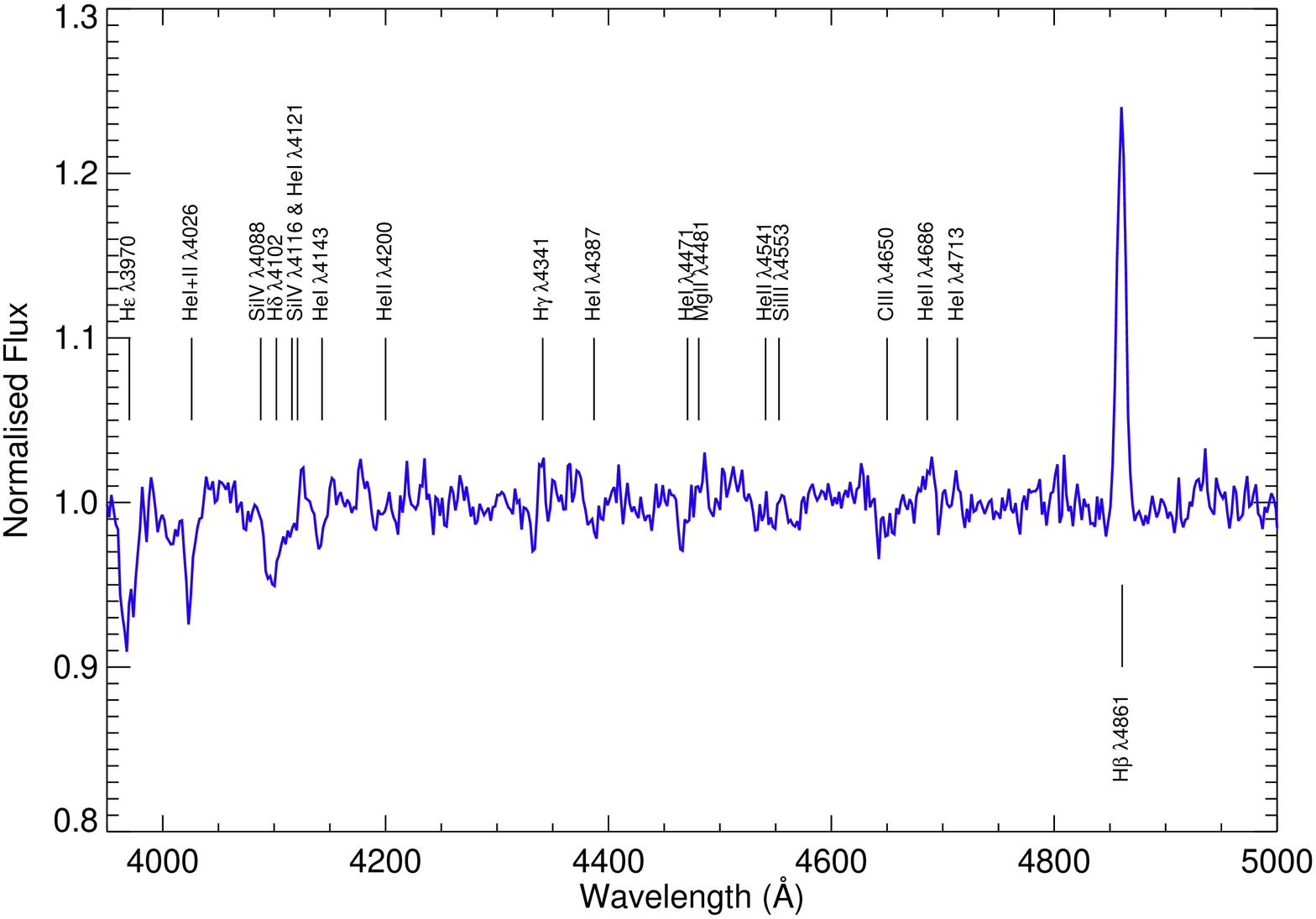}
\caption{Blue spectrum of \maxi\ obtained with the NTT on 2011 December 8. The spectrum has been normalized to remove the continuum and redshift corrected by $-158$~km~s$^{-1}$. Transitions relevant to spectral classification are marked. \\}
\label{fig:blue}
\end{figure*}

Figure \ref{fig:blue} shows the unsmoothed optical spectrum of \maxi, which is dominated by the Balmer and neutral helium lines. The He \textsc{ii} $\lambda\lambda$4200, 4143 lines are either too weak or absent, which makes the optical counterpart later than O9. The Si \textsc{iii}$\lambda$4553 line is stronger than Mg \textsc{ii} $\lambda$4481, implying a spectral classification earlier than B2.5. There does appear to be some evidence for the Si \textsc{iv} $\lambda$4116 line, however the rotationally broadened H$\delta$ line in close proximity makes it difficult to distinguish. As such we classify the optical counterpart of \maxi\ as a B1-2 star. 
We also compared our SAAO optical spectrum with those of classical Be stars obtained from the Be Stars Spectra Database (BeSS; \citealt{2011AJ....142..149N}), and \maxi\ is very similar to several of them (e.g., HD 277707 and HD 266894, see inset in Figure \ref{fig:optical_wide}). This further confirms the nature of the optical counterpart of \maxi. However, it is worth noting that there are no strong P-Cygni profiles observed in the emission lines, which might be expected for a major outburst event, but this could be attributed to the dominant continuum emission from the Be giant, thereby greatly diluting any such effect. 
As to its luminosity class, a B1-2 main-sequence star would have $M_V$ of $-3.2$ to $-2.5$. However, the lowest V measurement in Figure \ref{fig:lc} is 14.9 (corrected for the Galactic extinction), which corresponds to $M_V$=-4.0 (for $d$=60.6kpc), which is closer to that of a giant and the other two known Be+WD systems in the MCs, (XMMU J052016.0−692505 (LMC): $M_V$=-3.70, \citep{Kahabka06} with $\mu=-18.5$~mag, \citep{2008AJ....135..112S}; and XMMU J010147.5-715550 (SMC): $M_V$=-4.44, \citep{sssinsmc}). This is also typical of many other BeX systems in the SMC, which are in the range of $M_V=-5$ to $-2$ \citep{2012MNRAS.423.3663B}. 
We caution that the luminosity classification is very tentative as the reddening as well as the contribution of the circumstellar disk to the optical counterpart are poorly known. 

\subsection{ATCA Observations}
\maxi\ was observed by ATCA on 2011 December 23. 
The observations were performed simultaneously at frequencies 5.5 GHz and 9 GHz with configuration 6A (with the baseline ranging from 337 m to 6 km) and the upgraded Compact Array Broadband Backend. 
The data were taken with the CFB 1M-0.5k correlator configuration with 2 GHz bandwidth and 2048 channels, each with 1 MHz resolution.
The primary calibrator used was 1934-638, while the phase calibrator was 0230-790.
At the start of each observation, we observed 1934-638 for 10 minutes, and the phase calibrator every 25 minutes.
We used \texttt{MIRIAD} \citep*{1995ASPC...77..433S} to analyze the data with standard procedures.
We then performed standard data reduction steps, including bandpass, phase and amplitude calibrations.
When producing the dirty maps, we used the multi-frequency synthesis method \citep*{1999ASPC..180..419S} and natural weighting to suppress the noise.
The effective on-source integration time of \maxi\ was 3.8 and 3.4 hr at 5.5 GHz and 9 GHz, respectively. 
The field of view of ATCA with configuration 6A is $\sim$ 10$^{\prime}$ and 5$^{\prime}$ for 5.5 GHz and 9 GHz, respectively, and with spatial resolution that can reach $\sim$1$^{\prime\prime}$-- 2$^{\prime\prime}$. 

\subsubsection{ATCA Results}

\begin{figure*}[t]
\center
\psfig{figure=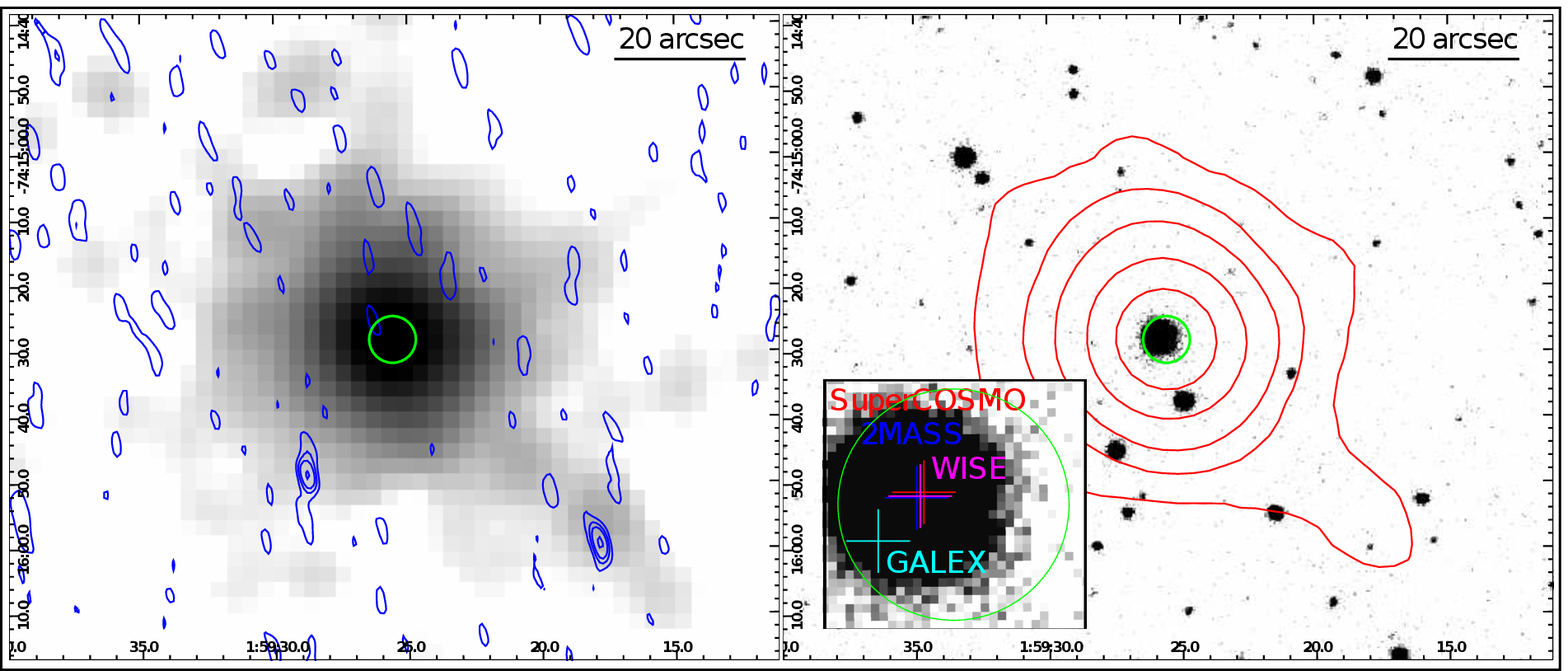,width=165mm}
\caption{Left: 1$\sigma$, 2$\sigma$, and 3$\sigma$ contours (blue) of the ATCA data are shown with a smoothed 25~ks \textit{Swift}/XRT (0.2--2 keV) image. The green circle indicates the XRT position of \maxi\ ($\mathrm{R.A.}=01$:59:25.59, $\mathrm{decl.}=-$74:15:28.21, J2000.0) with uncertainty $3\arcsec.6$ estimated by \texttt{xrtcentroid}. Within the error circle, no significant radio source was detected, but a very marginal signal ( $<2\sigma$) was seen at the edge of the circle. Right: \textit{Swift}/XRT contours (red) shown with an \textit{I}-band image taken by the 1.3~m Warsaw Telescope at LCO (OGLE) in 2010 July (in the quiescent state). The zoom-in image at the lower left corner indicates the positions of the archival measurements. Both images have the same field of view. \\}
\label{fig:xrt}
\end{figure*}

Figure \ref{fig:xrt} shows the \textit{Swift} X-ray images overlaid with ATCA radio map contours of \maxi. 
Besides background quasars, we did not detect any significant ($>$3$\sigma$) radio sources at the \maxi\ location. 
However, there is a very marginal signal ($\sim$2$\sigma$) detected within the XRT error circle, which will require deeper exposures to confirm. 
In naturally weighted maps, the rms noise level for \maxi\ is 16 (5.5 GHz) and 25 (9 GHz) $\mu$Jy$\,$beam$^{-1}$. 
We combined the observations from different frequencies in order to reduce the rms noise level at 6.8 GHz, but the improvement is not significant due to the short integration time.
As a result, the observations reached an rms noise level of 15 $\mu$Jy$\,$beam$^{-1}$, giving a 3$\sigma$ upper limit of $\sim$45 $\mu$Jy$\,$beam$^{-1}$ for \maxi. 
Although we did not find any statistically significant radio emission at the X-ray position, there is a 9.3$\sigma$ detection ($\sim$150$\mu$Jy$\,$beam$^{-1}$) at 5.5 GHz approximately 25$\arcsec$ away, while this source was not detected at 9 GHz.
We further calculated a lower limit of 0.54 to its spectral index ($\alpha$) by using the flux density at 5.5 GHz and the rms at 9 GHz. This is consistent with a background active galactic nucleus \citep{2005MNRAS.356..568N,2006A&A...451..457T}.

\subsection{Archival Measurements}
There is an optical/IR source listed in the SuperCOSMOS \citep{supercosmos2} and Two Micro All Sky Survey (2MASS; \citealt{2mass}) catalogs that has been associated with \maxi. Quoted magnitudes are $B_\mathrm{J}=15.05$, $R_{2}=14.898$,\footnote{For $R$-band photometry, we used $R_2$ in preference to $R_1$ since the former has higher signal to noise and better calibration.} $I=14.885$,$ J=14.8$, $H=14.8$, and $K=14.4$. 
The B-band value is consistent with our UVOT low-state measurements.
The 2MASS values are consistent with measurements from a quite separate IR study of the Magellanic Bridge, where the quoted values are $J=14.78\pm0.01$, $H=14.65\pm0.02$, and $K_\mathrm{s}$=$14.38\pm0.02$ \citep{herbig}. 
Furthermore, we also found counterparts of \maxi\ in the \textit{GALEX} catalog (GR6) (FUV: 14.66$\pm$0.02 and NUV: 14.64$\pm$0.01) and the \textit{WISE} catalog ($w1=14.02\pm0.03$, $w2=13.74\pm0.03$, $w3=12.54\pm0.29$, and $w4 >9.78$). 

\subsubsection{Spectral Energy Distribution}

\begin{figure}[t]
\includegraphics[width=84mm]{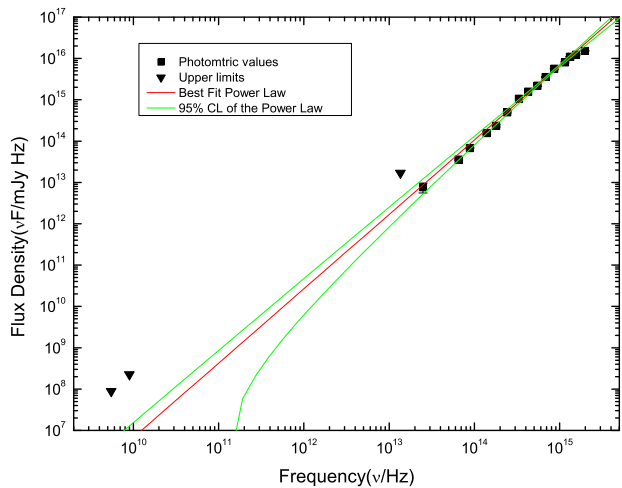}
\caption{
\maxi\ SED obtained by combining the ATCA radio data (upper limits at 5.5 GHz and 9 GHz); \textit{w1} (3.4$\mu$m), \textit{w2} (4.6$\mu$m), \textit{w3} (12$\mu$m), \textit{w4} (22$\mu$m) from the \textit{WISE} catalog; \textit{I} and \textit{R} from SuperCOSMOS; \textit{UVW1}, \textit{UVM2}, \textit{UVW2}, \textit{U}, \textit{B}, and \textit{V} from \textit{Swift} UVOT; and NUV, FUV from the \textit{GALEX}. The SED can be described by a power law of index about 1.8. \\}
\label{fig:sed}
\end{figure}

Using all the quiescent photometric measurements plus the two ATCA upper limits, we produced a spectral energy distribution (SED) from 5.5 GHz (5.45 cm) to 155 PHz (1928~\AA{}) (Figure \ref{fig:sed}). 
After applying extinction corrections, the SED can be represented as a power law of index $1.8$. 
According to \cite{1975A&A....39....1P}, the low frequency spectrum of the free--free radiation from a highly evolved star's extended envelope can be described as $\mathrm{F_\nu}\sim\nu^{0.6}$ (or $\mathrm{\nu\,F_\nu}\sim\nu^{1.6}$), which is consistent with the spectral index of our SED. Although \cite{1975A&A....39....1P} derived the $\nu^{0.6}$ relation for radio frequencies, they also stated that the relation holds over a wider range of frequencies as long as the radius of the envelope is sufficiently large. Furthermore, it is known that the mid-IR spectra of many Be stars are dominated by free--free emission \citep{1999AJ....118.2974R}, and this might explain the quiescent SED of \maxi. 
However, we caution that the extinction corrections may not be fully applicable in this region (we used the SMC Bar extinction curves), and hence the SED shape is only approximate. 

\section{Discussion}

\subsection{Nature of the Compact Object and Mass Donor in \maxi}

The sequence of \textit{Swift}/XRT spectra obtained throughout the outburst of \maxi\ have properties that are typical of classical SSS in the Magellanic Clouds, principally low-temperature ($\sim$100eV) thermal X-ray emission (after the initial flash) that is likely the consequence of thermonuclear burning of hydrogen on the surface of a WD, in stark contrast to the properties of BH/NS systems, where the inferred radii and temperatures are $R_\mathrm{in}\sim$10--100~km and $T_\mathrm{in}\sim$1~keV. 
Furthermore, our optical spectra are at velocities consistent with \maxi\ being located in the Magellanic Bridge; and the spectral absorption features later in the decline identify the mass donor as an early-type star, likely B1-2IIIe. 
Classical SSS like Cal83 and Cal87 are ``persistent" sources while \maxi\ is transient, and with a very short duration in both X-rays and optical. 
It also produced a strong ``hard" X-ray flash before entering the SSS phase, which makes it a very special case that cannot be explained easily within the standard SSS model. 
Combined with the inferred blackbody radius of the SSS  component (about $0.001$--$0.01$~$R_\sun$), we therefore conclude that \maxi\ is a WD orbiting a Be star. 

There is now a consensus on the accretion rate regimes and masses in which persistent and transient SSS occur, and this has recently been reviewed by \cite{Kato2010}. This can be summarized briefly as follows:

\begin{itemize}
 \item {\it $10^{-7}$--$4\times10^{-7} M_\sun\,$yr$^{-1}$}, in this relatively narrow range of mass transfer rates, accreting WDs (of mass 0.7--1.2 $M_\sun\,$) burn hydrogen on their surfaces steadily without driving substantial outflows, and are seen as persistent SSS;
 \item {\it below $10^{-7} M_\sun\,$yr$^{-1}$}, the hydrogen burning is triggered sporadically, which is normally seen as a nova explosion (either classical or recurrent). During the later phases of the nova outburst, the system will enter an SSS phase, the duration of which is a strong function of the WD mass, with the highest masses producing the shortest durations (see \cite{Kato2010}, and references therein);
 \item {\it above $4\times10^{-7} M_\sun\,$yr$^{-1}$}, the hydrogen burning drives substantial outflows which is seen as an optically thick wind. However, this wind can interact with the low mass donor in ways that can temporarily stop the Roche lobe overflow, which will then cause the SSS to switch off. Subsequently, the mass transfer onto the WD restarts and the process is repeated. The prototype of this behavior is RXJ0513.9-6951 \citep{Southwell96}.
\end{itemize}

Based on this interpretation, \maxi\ fits the second (low-mass transfer rate) explanation extremely well, and so we can constrain the mass accretion rate onto the WD in \maxi\ to be less than a few times $10^{-7} M_\sun\,$yr$^{-1}$. But what about the nova explosion that should have been associated with this event? In fact, typical novae in the SMC have peak magnitudes in the range 11--13.6,\footnote{See http://www.mpe.mpg.de/$\sim$m31novae/opt/smc/index.php. } but these are all classical or recurrent novae, where the mass donor is a low-mass star (and hence intrinsically faint). In the case of \maxi, the donor is intrinsically a luminous star in its own right, and so the approximately 0.5--1mag increase observed here (Figure \ref{fig:lc}) is entirely consistent with a classical or recurrent nova event taking place in the company of an early-type star.

That there are multiple emitting components in this system is clear from the different temporal behavior that is a strong function of wavelength. The X-ray/optical lightcurves (Figure \ref{fig:lc}) do show a clear positive correlation, where the X-ray emission drops exponentially for about two weeks and the \textit{U}-band light drops even faster (at a rate of 0.5 mag in 2.4 days). More significantly, the duration of the optical outburst is about six times shorter than that of the X-ray outburst. \maxi\ also shows significant excess (0.1--0.5 mag) in the other three longer wavelength (\textit{B}, \textit{V}, \textit{UW1}), yet no significant excess can be found at shorter wavelengths (\textit{UM2}, \textit{UW2}). 
This break in behavior at UV wavelengths, and the huge difference in the outburst durations indicates that the X-ray origin is intrinsically distinct from the optical, and hence infers the presence of multiple emitting regions. 

Radio detection should also be expected for such a nova-like scenario. However, such radio emission would be much too weak at the distance of the SMC for our ATCA observation ($>$10~$\mu$Jy rms noise level). For examples, V Sge radiated 70~$\mu$Jy at 1.2--2.75~kpc, equivalent to 6$\times10^{-4} \mu$Jy at 60~kpc) \citep{1997MNRAS.287L..14L}, and SS Cyg with an accretion rate of around $10^{-8} M_\sun\,$yr$^{-1}$ radiated $\sim$0.2~mJy at 100~pc (equivalent to 0.15~$\mu$Jy at 60~kpc) \citep{2011MNRAS.418L.129K}. Also, the ATCA observation occurred $>$1 month after the outburst, when the expected radio emission would have been even fainter.

It is instructive to compare the X-ray light curves of \maxi\ and the recurrent nova RS Oph \citep{2011ApJ...727..124O}, as both systems showed dramatic variations (up to one order of magnitude) on short timescales ($10^3$--$10^4$~s). RS Oph showed a general trend that the X-rays are softer when the flux is stronger, which was also present in \maxi\ (Figure \ref{fig:hr}). 
We also compared \maxi\ with a classical nova (V2491 Cyg) that was well studied in X-ray and optical during its SSS phase \citep{2010MNRAS.401..121P}. 
Both SSS have similar lightcurves in X-ray and optical that decay exponentially following outburst. However, the optical amplitudes (6--7~mag) and durations ($\sim$150 days) of V2491 Cyg are much higher and longer than those of \maxi\ (0.5~mag in 1--2 days). 
However, the fundamental difference between \maxi\ and classical/recurrent novae is the nature of the mass donor. Accordingly, we argue that the optical decline phase of \maxi\ is largely hidden by glare of the Be primary, otherwise the magnitude change and duration would have been much higher and longer. A similar effect has been observed in the case of the \textit{Fermi}-detected nova V407 Cyg \citep{2012ApJ...748...43N}. 
It is also very likely that the optical peak of \maxi\ was missed, so that we have underestimated the maximum brightness of \maxi. 
Moreover, the \textit{I}-band flux (OGLE-IV) of \maxi\ was above the quiescent level for at least 60 days after the major flare. This means a much higher \textit{V}-band duration is highly possible but just hidden by the limitations of \textit{Swift}/UVOT. 

The brevity of the SSS phase ($\leq$15 days), together with the high temperatures ($\sim$90-110eV) from both blackbody and WD atmosphere models, provides an indication that the WD in \maxi\ is massive. According to \cite{Nomoto07} such temperatures are only reached in white dwarfs exceeding 1.3 $M_\sun$. Furthermore, the SSS light curve of \maxi\ is very similar to that of the very fast classical nova V2491 Cyg, whose SSS phase was only 10 days. This led \cite{HachisuKato09} to infer a WD mass of $\sim$1.3 $M_\sun$.

\subsection{Origin of the X-Ray Spectral Features}

Early in the outburst, the \textit{Swift}/XRT spectra (Figure \ref{fig:spec}) required additional spectral features (a broad emission line at ${\sim}0.7$~keV and an absorption edge at $0.89$~keV). These could be associated with O, Fe, and Ne features, where the broad 0.7~keV emission is a combination of \oviii\ Ly$\alpha$ (650~eV), Fe L (700--720~eV), and Ne K (850~eV) lines; and the 0.89~keV absorption edge could be associated with the K-edge of \oviii (0.88~keV). 
A similar absorption edge has been observed in another SSS, RXJ0925.7-4758, which is undergoing steady nuclear burning on the WD surface \citep{sssedge}. 
Consequently, the K-edge and 0.7~keV broad emission features also support the interpretation of \maxi\ as an accreting WD system. Their origin is likely to be in the ejected shell from the WD surface at the beginning of the strong outburst, following which the WD enters a luminous supersoft phase \citep{2003MNRAS.341L..35K}. This provides further support for \maxi\ containing a massive ($\geq$1.3$M_\sun\,$) WD. 
If it is true, the accreting WD in the system could be O-rich. 

\subsection{Origin of the Initial X-Ray Flash}

So far we have concentrated on the analysis and interpretation of the \textit{Swift}/XRT and UV/optical observations of \maxi. However, perhaps the most remarkable feature of this outburst is the initial X-ray flash detected by the MAXI/GSC. Having established now that \maxi\ is in the SMC, then the luminosity of this flash, assuming a $\sim$1~keV thermal (Raymond-Smith) plasma model, is ${\sim}1.6\times10^{39}(d/60.6\,\mathrm{kpc})^2$\lum\ (2--4~keV). During that single orbit, \maxi\ outshone the luminous NS X-ray binary, SMC X-1, in the MAXI/GSC 2--4~keV energy band. Even more remarkably, the flash was very short-lived, rising and falling in $<$92 minutes by at least a factor of 20, as MAXI detected no source at that location in the orbits before and after the X-ray flash.

The first \textit{Swift} observation was approximately 10 hr later, and revealed the lower temperature SSS component, which we believe is quite distinct from the harder X-ray flash. That is because, even though the SSS flux declined over subsequent days, the timescale for this was much longer than the dramatic drop of the X-ray flash. Consequently, we should consider alternative emission mechanisms for the X-ray flash.

First, we have now argued that many of the \maxi\ properties are accountable as a nova explosion, and these can lead to hard X-ray components being present, as well as a short-lived SSS phase (see e.g., \cite{Mukai08}, and references therein). An excellent example of this is the recurrent nova, RS Oph, that underwent a well-studied outburst in 2006 \citep{Sokoloski06}, reaching a hard X-ray luminosity of $\sim$10$^{36}$\lum, which included Fe K$\alpha$ emission. This was explained as the result of a $\sim$3500\kms\ ejected nova shell hitting the wind of the M giant donor star, essentially a ``mini-SNR'' as the subsequent X-ray decline followed a typical Sedov light curve. However, in some cases, the X-ray properties imply that the emission originates internal to the nova ejecta, and so internal shock models have been developed \citep{MukaiIshida01}. Nevertheless, the X-ray luminosities attained are usually $\leq$10$^{35}$\lum. 

But much higher luminosities are possible in special circumstances. The 1998 X-ray transient, XTE J0421+560, was associated with the bright B[e] star, CI Cam, and initially interpreted as the periastron passage of an NS or a BH compact object through the dense equatorial outflow from the B[e] star \citep{Hynes2002}, producing a peak X-ray luminosity of ${\sim}3\times10^{37}(d/2\,\mathrm{kpc})^2$\lum\ (3--20~keV). However, CI Cam has properties that did not fit well with other high-mass X-ray binaries or soft X-ray transients, and this led \cite{Filippova2008} to propose that XTE J0421+560 was actually a nova explosion on an accreting WD in its 19 days, eccentric orbit. While this interpretation, and the nature of the compact object in CI Cam, remains controversial, the detailed analysis by \cite{Filippova2008} of a nova explosion in the surroundings of an early-type star remain directly applicable to \maxi. The large increase in luminosity compared to RS Oph they explain as due to the much denser medium that 
surrounds the B[e] star. The ejected nova shell interacts with and shocks the B[e] wind creating a hot ($\sim$1--10~keV) thermal plasma that both heats and cools rapidly. In the case of XTE J0421+650 there was sufficient coverage to follow the rise and 
fall of this flux by a factor of 10 in $<$1 day \citep{Filippova2008}. This is possible given a dense B[e] wind ($\sim$10$^{10}$cm$^{-3}$), which the shock will increase by a factor of four, and will radiatively cool on the observed timescale. 

We therefore consider whether the detailed calculations of \cite{Filippova2008} can be applied to \maxi, and what the implied parameters of the system would be. They give the timescale for radiative cooling as ${\sim}3kT/2{\Lambda}n$s, where $\Lambda$ is the emissivity of a hot gas, and is $\sim$10$^{-24}$erg~cm$^{3}$s$^{-1}$ in the 2--20keV band for $kT\sim$1keV. The \maxi\ X-ray flash cooled in $\sim$4000s, and this requires $n\sim$5$\times$10$^{11}$cm$^{-3}$. Allowing for the factor of four increase by the shock, this is only a factor of 10 higher than that inferred for CI Cam by \cite{Filippova2008}, and entirely reasonable for the enhanced density in the equatorial outflow of a rapidly rotating Be star.

Unfortunately, we do not yet know the orbital period of the \maxi\ system, but we assume 10d in order to make it more compact, and hence a higher density circumbinary medium. Taking the B1/2IIIe donor as $\sim$10$M_\sun\,$ and a massive (1.3$M_\sun\,$) WD, then a 10 days period gives an orbital separation, $a{\sim}$1.5$\times$10$^{13}$cm. We can then use the order of magnitude calculation of the resulting shocked X-ray luminosity, $L_\mathrm{CBM}$, from \cite{ItohHachisu90}:

\begin{equation}
\begin{split}
 L_\mathrm{CBM} = 3{\times}10^{33} \frac{\Lambda}{10^{-23}\mathrm{erg~cm^3s}^{-1}} {\frac{M_\mathrm{CBM}}{10^{-6}M_\sun}}^2 \\ \times{\frac{r_\mathrm{CBM}}{10^{15}\mathrm{cm}}}^{-2} {\frac{r_s}{10^{15}\mathrm{cm}}}^{-1} \mathrm{\,erg\,s}^{-1}, 
 \end{split}
\end{equation}

where $\mathrm{CBM}$ refers to ``circumbinary medium'', and $r_s$ is the radius of the shock wave. Taking both $r_s$ and $r_\mathrm{CBM}{\sim}a$, and $M_\mathrm{CBM}\sim$10$^{-6}M_\sun\,$ (at least this much is normally ejected in a nova explosion), then $L_\mathrm{CBM}\sim$10$^{39}$\lum. This is an order of magnitude or so above that of CI Cam, but the BeX in the SMC have well established, dense equatorial outflows as a result of their rapid rotation \citep{2005MNRAS.356..502C,2007ApJ...660..687M} and so these parameters are by no means extreme. 
In addition to accounting for the X-ray flash luminosity, the model also predicts that the SSS turn-on can be seen almost instantly, as the X-ray flash ionizes the ejected shell.
Furthermore, if the compact object is a massive O--Ne--Mg WD, then enhanced X-ray emission features at the energies seen in our \textit{Swift}/XRT spectra would be expected.

\subsection{Be+WD Systems in the Magellanic Clouds}

There have already been two SSS discovered that appear to be associated with Be mass donors, the first, XMMU J052016.0-692505, in the LMC \citep{Kahabka06}, the second, XMMU J010147.5-715550, in the SMC \citep{sssinsmc}. The LMC SSS has $L_X>$10$^{34}$\lum, whereas the SMC SSS is weaker, at 3$\times$10$^{34}$\lum. Both are optically identified with Be systems, and appear to be persistent SSS and variable. Consequently, they are both likely in the steady hydrogen burning phase, and neither have exhibited nova eruptions comparable to \maxi. Interestingly, both have reported very long periodicities in their OGLE light curves ($\sim$510 or 1020 days, and 1264 days for the LMC and SMC sources, respectively), although these are not well constrained due to the limited monitoring timescale. However, such long-term periodicities are now a well-established feature in SMC BeX sources \citep{Rajoelimanana}, in many of which the much shorter orbital periods are already well known. Consequently, we dismiss the long periods in these new SSS from being orbital in nature. And in any case, the orbital periods will have to be very much shorter in order to sustain the mass accretion necessary for steady hydrogen burning, as calculated in the previous section.

\subsection{Comparison with QSS/SSS in Nearby Galaxies}

Over the last decade, many large extragalactic surveys have been undertaken by \textit{Chandra} and \textit{XMM-Newton} in order to search for SSS and possibly related systems (so-called quasi-soft sources, or QSS) in nearby galaxies of the Local Group \citep{d2003,d2004,2004ApJ...610..247D,kong2005}. Some of the candidates discovered in these surveys have also reached ``ultraluminous'' levels, as well as displaying transient behavior. 
One of the best examples is ULX-1 in M101 \citep{kong2005}. 
Apart from the much higher implied luminosity, \maxi\ is quite similar to M101 ULX-1 in many aspects (i.e., temperature, variability). Interestingly, M101 ULX-1 also has a massive companion, a B supergiant counterpart \citep{2005ApJ...620L..31K}, close to a star formation region. Consequently, it is worth considering whether these are all examples of the same phenomenon.

A popular, but controversial, interpretation of extragalactic QSS and SSS is that they are extreme versions of the stellar-mass BH systems in our Galaxy, wherein the compact object is a much higher, IMBH, $\geq$50$M_\sun\,$ in a soft/high accretion state \citep{2010AN....331..205D}. However, by considering the similarities between these candidate IMBHs and \maxi, we propose that some of the IMBH candidates could instead be a \maxi -like object, consisting of a WD accreting from a high-mass donor and undergoing what are essentially classical or recurrent nova outbursts in a high-density environment. 
According to the work of \cite{d2004}, some QSS/SSS may have high-mass companions, based on their association with their host galaxy's spiral arms. Unfortunately, the main focus of this work was still on the IMBH scenario and the nature of the high-mass companion was not considered in detail. 

In fact, more than 40 QSS/SSS candidates have been discovered in M31 by \textit{Chandra} \citep{2004ApJ...610..247D}. 
All of them have low blackbody temperatures, and some of them exhibit a harder power-law component in their spectra. 
After comparing them with our QSS/SSS candidate \maxi, we propose that this kind of spectrum could be fully explained by our Be+WD binary model, where the low-temperature blackbody is indeed an accreting WD and the hard power-law component can be attributed to the shock interaction of the nova shell/ejection with the hot stellar wind (or Be equatorial, outflowing disc). Unfortunately, the even greater distances of this population of QSS/SSS make it very difficult to distinguish between various combinations of spectral models. It is therefore essential to obtain much higher quality spectra of these extragalactic systems, so as to probe the true nature of these supposedly IMBH systems.

\section{Conclusions}

By considering its WD-like radius inferred from the \textit{Swift}/XRT X-ray spectra ($0.001$--$0.01$~$R_\sun$) and its B1/2IIIe spectral type deduced from the ESO optical spectrum, we believe that \maxi\ is a member of the long-sought population of Be--WD binaries, and the first example to possibly enter ULX territory. 
To account for its properties (very short SSS phase of $\leq$15 days) we deduce that \maxi\ is a heavy ($\sim$1.35~$M_\sun$) O--Ne WD which is slowly accreting material from the wind of its early-type companion until it undergoes unstable hydrogen burning on the WD surface, in what is essentially a classical nova explosion. 

Furthermore, with its location in the SMC, \maxi\ is a key example of the extreme behavior (apparently ultra-luminous X-ray flash followed by SSS phase) that a nova explosion can lead to if it is in the appropriate environment. 
It has pointed us in a new direction for understanding the QSS/SSS in external galaxies that have been proposed as IMBH associated with massive companions. If confirmed, it reveals an entirely new sub-class of SSS, showing once again that WDs are capable of mimicking BHs, and possibly even (some) transient ULXs.

\begin{acknowledgements}
One of us (PAC) thanks Mike Bode, Koji Mukai, and Valerio Ribeiro for stimulating discussions on the nature and properties of nova outbursts. 
We also acknowledge the project ``Revealing the nature of southern hard X-ray sources through optical spectroscopy" for the SAAO optical spectrum presented here. 
This project is supported by the National Science Council of the
Republic of China (Taiwan) through grants NSC100-2628-M-007-002-MY3
and NSC100-2923-M-007-001-MY3. 
This research has made use of the MAXI data provided by RIKEN, JAXA, and the MAXI team.
This work is based on observations obtained with \textit{Swift}, a part of NASA's medium Explorer program, with the hardware being developed by an international team from the United States, the United Kingdom, and Italy, with additional scientific involvement by France, Japan, Germany, Denmark, Spain, and South Africa.
The OGLE project has received funding from the European Research Council
under the European Community's Seventh Framework Programme
(FP7/2007-2013)/ERC grant agreement No. 246678 to AU. 
We thank the ATCA team for an allocation of Director's Discretionary Time for this project. 
The Australia Telescope is funded by the Commonwealth of Australia for operation as a National Facility managed by CSIRO. 
This publication also makes use of data products from the \textit{Wide-field Infrared Survey Explore}r, which is a joint project of the University of California, Los Angeles, and the Jet Propulsion Laboratory/California Institute of Technology, funded by the National Aeronautics and Space Administration.

\end{acknowledgements}

\bibliographystyle{apj}
\bibliography{MAXI}
\end{document}